\newif\ifarxiv
\arxivtrue

\documentclass{PoS}

\usepackage{amssymb}
\usepackage[loose]{units}

\usepackage[comma,square,numbers,sort&compress]{natbib}
\newcommand{\xt}           {\ensuremath{x_{\mathrm{T}}}}
\newcommand{\pt}           {\ensuremath{p_{\mathrm{T}}}}
\newcommand{\kt}           {\ensuremath{k_{\mathrm{T}}}}
\newcommand{\xitg}           {\ensuremath{\xi_{\mathrm{T}}^\gamma}}

\newcommand{\Ncoll}        {\ensuremath{N_{\mathrm{coll}}}}
\newcommand{\Npart}        {\ensuremath{N_{\mathrm{part}}}}

\newcommand{\Nch}        {\ensuremath{N_{\mathrm{ch}}}}
\newcommand{\Rg}        {\ensuremath{R_{\gamma}}}
\newcommand{\RAA}        {\ensuremath{R_{\mathrm{AA}}}}
\newcommand{\NAA}        {\ensuremath{N_{\mathrm{AA}}}}
\newcommand{\ZAA}        {\ensuremath{Z_{\mathrm{AA}}}}
\newcommand{\sipp}       {\ensuremath{\sigma_{\mathrm{pp}}}}
\newcommand{\spp}        {\ensuremath{\sqrt{s}}}
\newcommand{\sNN}        {\ensuremath{\sqrt{s_{\mathrm{NN}}}}}
\newcommand{\Mee}        {\ensuremath{M_{\mathrm{ee}}}}
\newcommand{\Mmm}        {\ensuremath{M_{\mathrm{\mu\mu}}}}
\title{Shining a Light on the QGP - \\ Electroweak Probes Experimental Summary}

\ShortTitle{Electroweak Probes Experimental Summary}

\author{\speaker{Friederike Bock}\\
        CERN\\
        E-mail: \email{friederike.bock@cern.ch}}

\abstract{Objects which are only subject to the electroweak force are an ideal probe of QCD in high density and temperature environments as they carry information about the conditions during their production out of the QGP without interacting with it. Thus they can be used to characterize the initial state as well as several properties of the QGP and its evolution. Within this article the recent results regarding these electroweak probes presented at the 9$^{th}$ Hard Probes Conference (Aix-les-Bains, France, 2018) will be summarized. In particular the following questions will be addressed: Can we determine the necessary energy and particle density for the QGP creation using electromagnetic probes? Can we use light-by-light scattering in heavy-ion collisions to do precision tests of QED and measure the magnetic field of the QGP? How are quark jets modified by the presence of the QGP? What can we learn about the initial state and scaling properties in pp, p(d)-A and A-A collisions from the production of high \pt\ photons, Z$^0$ and W$^\pm$ bosons? \ifarxiv \\ In the version provided on arXiv, additional plots are included in the appendix with respect to the version submitted for publication in the journal.\fi}

\FullConference{International Conference on Hard and Electromagnetic Probes of High-Energy Nuclear Collisions\\
		30 September - 5 October 2018\\
		Aix-Les-Bains, Savoie, France}

\begin{document}

\section{Introduction}
\vspace{-0.3cm}
Ultrarelativistic heavy-ion collisions provide the perfect laboratory to study QCD at high temperatures and high energy density. In these collisions a Quark-Gluon-Plasma (QGP) is created and subsequently thermalizes. The plasma cools down and expands, until it finally hadronizes and the hadrons stop interacting among themselves and free-stream. Electroweak probes offer a unique perspective on the properties of the QGP and the hot hadron gas (HG) as they do not interact strongly with the medium and are produced at all stages of the collision. Depending on their production mechanism the different probes allow access to different aspects of the heavy-ion collision.\\
In analogy to black body radiation the thermal medium created in heavy-ion collisions will emit electromagnetic radiation in form of low momentum photons and di-leptons. The measurement of the low-\pt\ (virtual) direct photon excess beyond the expectations from next-to-leading order (NLO) pQCD calculations is one of the important signatures for establishing a QGP creation. It has been found that the hadron production in high multiplicity p-p and p(d)-A collisions exhibits similar collective behaviors as seen in A-A collisions \cite{Loizides:2016tew}. The exact nature of this effect is still unclear, however. Measuring a significant direct photon signal and the corresponding effective temperature of the source should provide more information about the nature of the medium created in these collisions. Furthermore, the measurement of the di-lepton spectrum in A-A collisions can give insights to the in-medium electromagnetic spectral function of the $\rho^0$ and its possible modification due to chiral-symmetry restoration. \\
If instead the electroweak probes are produced in the initial hard scattering processes, like the weak bosons (W$^\pm$, Z$^0$) and most of the direct photons at high-\pt, they are ideal probes for the parton distribution functions of the incoming nuclei and protons. Additionally, their yield at mid-rapidity can be used to verify the Glauber model as their production should scale with the number of binary collisions. Moreover, they can be used as calibration probes for the recoiling jets and their in medium modifications, as they serve as a measure of the initial energy of the jet and allow to discriminate between quark and gluon induced jets.\\
Electromagnetic probes in heavy-ion collisions, however, do not only provide a stringent test of QCD, but even test QED under extreme conditions. Due to the large fluxes of quasireal photons emitted by the nuclei accelerated to TeV energies at large distances from the center of the nuclei light-by-light scattering can occur in heavy-ion collisions at RHIC and the LHC \cite{Bertulani:2005ru}. The first evidence for the presence of these processes in ultraperipheral Pb-Pb collisions at the LHC has been presented by the ATLAS collaboration \cite{Aaboud:2017bwk}. Observing these processes and their modifications in more central collisions could help determining the magnetic field strength of the QGP itself \cite{Kharzeev:2009pj}.\\
Within this article a summary of the results on electroweak probes presented at the 9$^{th}$ Hard Probes Conference (Aix-les-Bains, France, 2018) will be given. Particular focus lies on addressing the following questions and the results which contributed to their answers. Each section serves as a summary on the corresponding topic: Can we determine the necessary energy and particle density for the QGP creation using electromagnetic probes (\href{sec:QGPProp}{Section \ref{sec:QGPProp}})? Can we use light-by-light scattering in heavy-ion collisions to do precision tests of QED and measure the magnetic field of the QGP (\href{sec:lbl}{Section \ref{sec:lbl}})? How are quark jets modified by the presence of the QGP (\href{sec:probes}{Section \ref{sec:probes}})? What can we learn about the initial state and scaling properties in pp, p(d)-A and A-A collisions from the production of high \pt\ photons, Z$^0$ and W$^\pm$ bosons (\href{sec:IS}{Section \ref{sec:IS}})? 

\section{Probing the QGP with Direct Photons and Di-Leptons}
\label{sec:QGPProp}
\begin{figure}
  \includegraphics[width=0.5\textwidth]{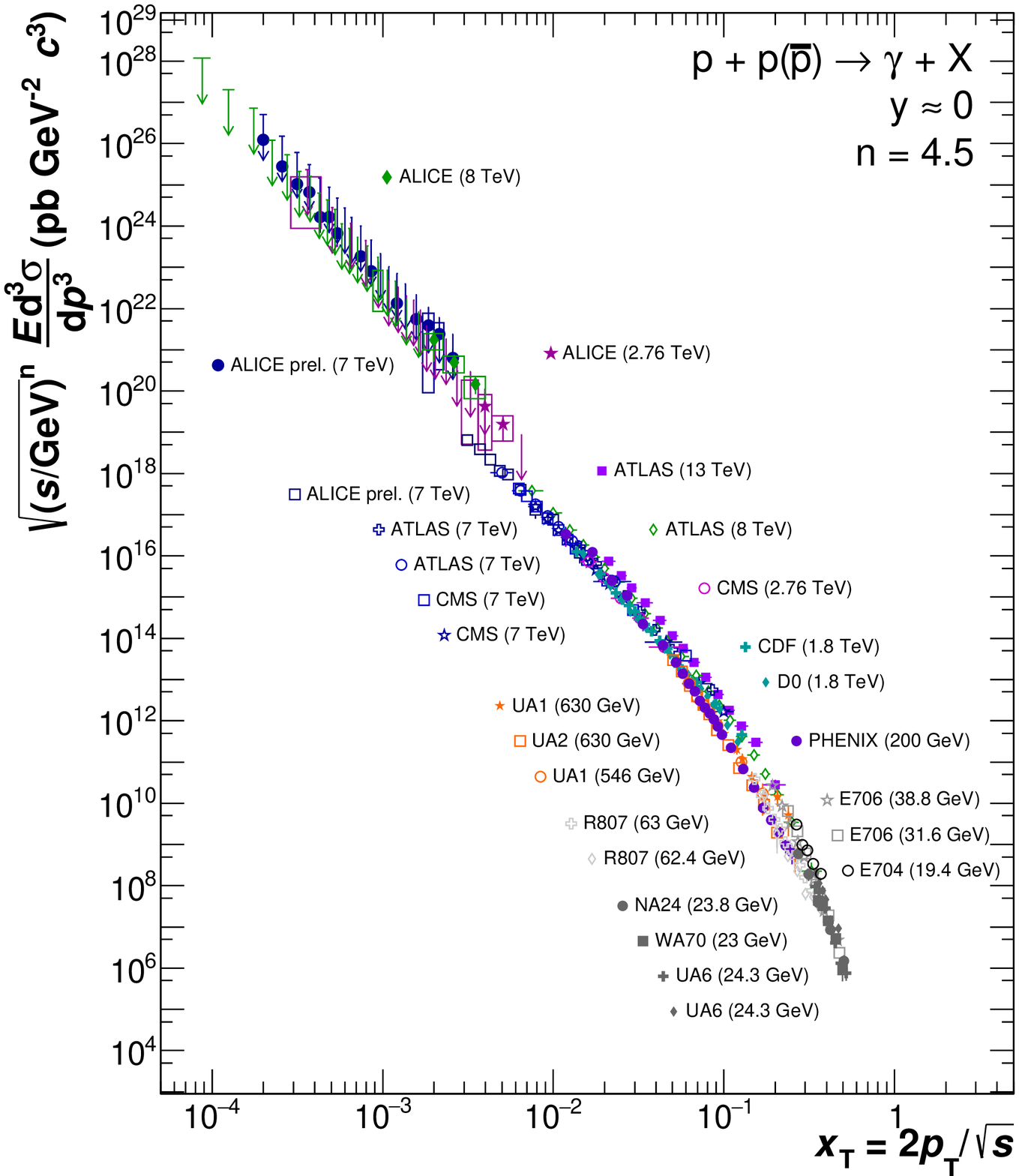} 
  \includegraphics[width=0.5\textwidth]{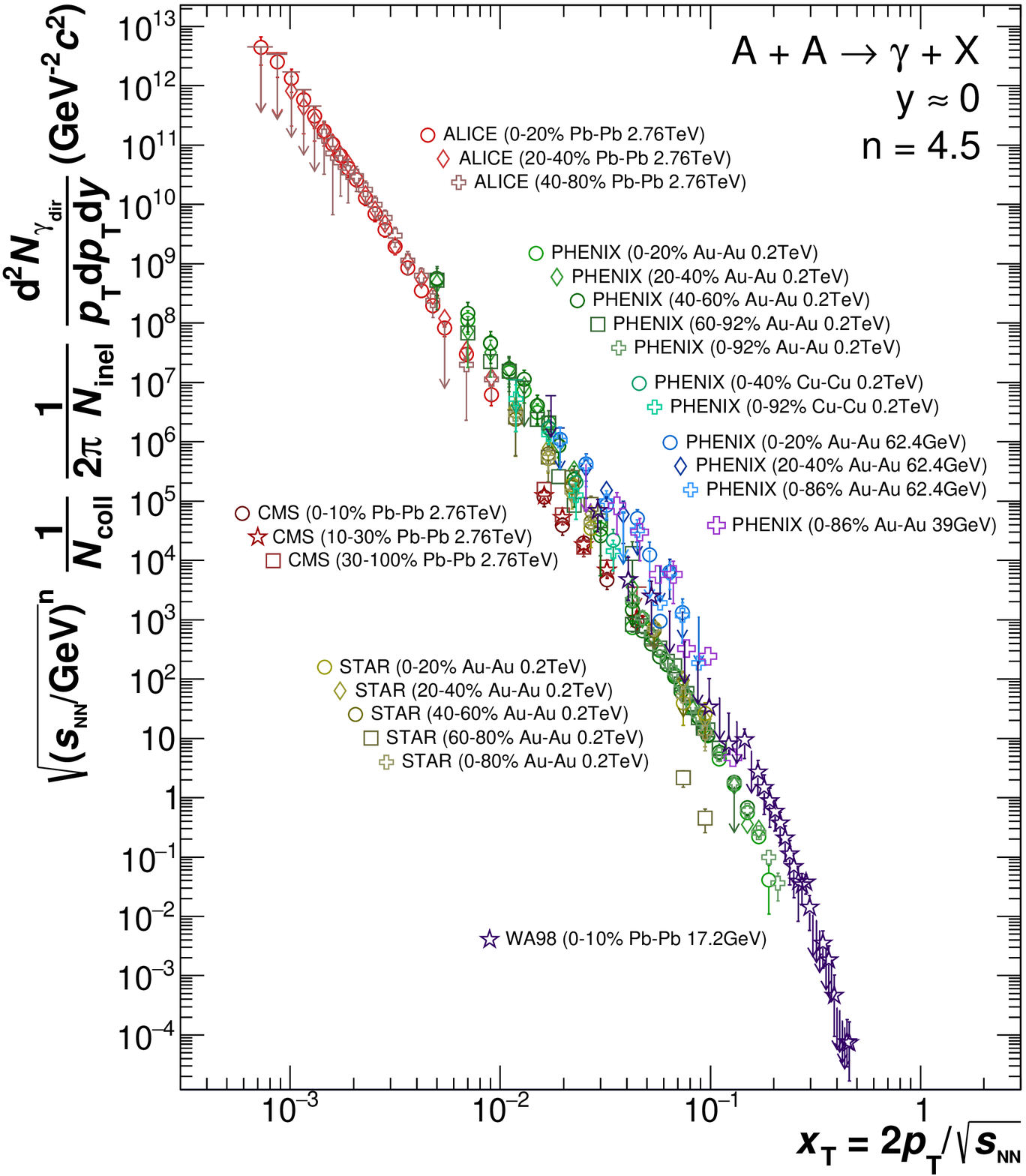} 
  \caption{Left: Summary of \xt-scaled (isolated) direct photon invariant cross sections measured in pp and p$\bar{\mathrm{p}}$ collisions at various collider energies 
  \ifarxiv
    \cite{Adare:2012yt,Abachi:1996qz,Abe:1994rra,Albajar:1988im,Albajar:1989sr,Alitti:1991yk,Adams:1995gg,Apanasevich:2005gs,Apanasevich:2004dr,
  Alverson:1993da,DeMarzo:1986vi,Bonesini:1987bv,Sozzi:1993sm,Ballocchi:1998au,Aad:2011tw,Aad:2010sp,Aad:2016xcr,Aaboud:2017cbm,Wilde:2012wc,Acharya:2018dqe,Chatrchyan:2011ue,Khachatryan:2010fm} 
  \else
    \cite{Adare:2012yt,Aad:2011tw,Aad:2016xcr,Aaboud:2017cbm,Wilde:2012wc,Acharya:2018dqe,Chatrchyan:2011ue}
  \fi
  scaled with a common $n=4.5$. Right: Summary of \Ncoll\ and \xt-scaled (isolated) direct photon invariant yields measured in A-A collisions at various collider energies \cite{Gale:2009qy,Aggarwal:2000th,STAR:2016use,Adare:2018wgc,Adare:2018jsz,Adare:2014fwh,Afanasiev:2012dg,Adare:2009qk,Adare:2008ab,Adler:2005ig,Aad:2015lcb,Chatrchyan:2012vq,Adam:2015lda} scaled with a common $n=4.5$ and the number of collisions (\Ncoll) per collision system and respective centrality.}
  \label{fig:xTppandAA}
\end{figure}
While it has been established that a QGP is created in central heavy-ion collisions at SPS, RHIC and LHC energies the question remains whether the particle and energy density is sufficient for it to be also created in peripheral A-A collisions or even smaller systems, like high multiplicity p-p or p-A collisions. 
The measurements of the low momentum (virtual) direct photons, which are presumed to be of thermal origin, served as one of the key pillars for establishing the creation of a QGP. 
These low \pt\ photons can be measured as an excess (\Rg) above the expected decay photon spectrum originating from $\pi^0$, $\eta$, $\omega$ and other mesons and baryons decaying into photons.
Furthermore, they can be extracted from the di-lepton invariant mass spectrum using the Kroll-Wada-equation, after subtraction of the resonance contributions.
Within the excess, however, not only thermal photons are contained, but also photons from hard scattering processes as well as fragmentation photons. 
Above a transverse momentum of about 4 GeV/$c$ the dominant source of photons are those from the later processes, which are well described by next-to-leading order pQCD calculations. 
Below those momenta, on the other hand, NLO pQCD calculations are not constrained any longer and the photon yield originating from the hard scatterings can only be extracted with large uncertainties. 
In order to reduce these uncertainties the experimental collaborations are pursuing the same measurement in pp collisions, where mainly the photons from the hard scattering and the fragmentation of jets should be present, which presumably scale with the number of collisions (\Ncoll) in a heavy-ion collision. \\
At this conference new results regarding the direct photon production in pp collisions at \spp$~=~0.2,$ $2.76,~7,~8$ and $13$~TeV have been presented by the PHENIX and ALICE collaborations \cite{Khachatryan:2018evz,Acharya:2018dqe,Acharya:2018ohw}.
While the PHENIX collaboration has been able to extract a direct photon spectrum down to about 1.5 GeV/c at \spp$= 0.2$ TeV the same measurements have proven to be challenging at the LHC due to the smaller fraction of direct photons with respect to the decay photons at the LHC. 
Thus the extraction of the direct photon spectra at LHC energies was only possible above $6$ GeV/c\ifarxiv~(\href{fig:DRppandpPb}{Figure \ref{fig:DRppandpPb}})\fi, where they are well described by pQCD calculations, and upper limits were obtained for lower transverse momenta. 
When taking into account all measured direct and isolated photon spectra at current and previous colliders \ifarxiv (\href{fig:dirGppandPbPb}{Figure \ref{fig:dirGppandPbPb} (left)}) \fi it has been found that the high \pt\ spectra exhibit \xt-scaling \cite{Adare:2012yt}. 
\href{fig:xTppandAA}{Figure \ref{fig:xTppandAA} (left)} shows the updated compilation of all direct and isolated photon spectra measured at mid-rapidity in pp or p$\bar{\mathrm{p}}$ collisions from \spp$~=0.019$ to $13$ TeV as a function of $\xt= 2\pt/\spp$. 
The effective scaling parameter $n$ was chosen to be $4.5$ as obtained in \cite{Adare:2012yt}. 
It can be seen that \xt-scaling seems to hold within $20-50\%$ if the spectra are scaled with $\spp^n$ and plotted as a function of \xt.
However, a discontinuity is seen for the spectra obtained at the LHC for the measurements from the ALICE collaboration using the $\Rg$-technique and the isolated direct photon spectra from ALICE, CMS \& ATLAS.
This discontinuity arises from the fact that the isolated photon spectra do not contain the fragmentation photons, while prompt photons cannot be discriminated from the fragmentation photons using the \Rg-technique.
Even though a full reanalysis of the obtained data regarding the effective scaling parameter should be performed, it seems that the spectra obtained a higher \spp\ can be used to constrain the low momentum spectra at lower \spp, yielding stronger constrains than the upper limits provided by the measurements for the NLO pQCD calculations.\\
In addition to the base line measurements in minimum bias pp collisions the PHENIX and ALICE collaborations presented new measurements regarding the low momentum direct photon production in p-A collisions as a function of centrality. \ifarxiv The compilation of the available measurements in p(d)-A collisions ranging from \sNN$~= 0.2 - 8.16$~TeV can be found in \href{fig:dirGpA}{Figure \ref{fig:dirGpA}}. \fi
The minimum bias measurements in p-A collisions show no significant excess of direct photons beyond the expected yield of prompt and fragmentation photons calculated using NLO pQCD \cite{Schmidt:2018ivl,Khachatryan:2018evz}. 
In 0-5\% central p-Au collisions on the other hand a slight increase of direct photons below $3$ GeV/$c$ has been found by the PHENIX collaboration, indicating that a thermal source might be present in these collisions \cite{Khachatryan:2018evz}. 
A similar analysis versus multiplicity in p-Pb collisions at $\sNN=5.02$ TeV, also showed a slight increase in the direct photon excess \ifarxiv(\href{fig:DRppandpPb}{Figure \ref{fig:DRppandpPb}}) \fi for the most central 0-20\% events.
Within the systematic and statistical uncertainties, however, it is not yet possible to claim a significant direct photon signal at this center of mass energy for $\pt~<~3$ GeV/c.\\
Increasing the system size even more the PHENIX collaboration also showed new results on the direct photon production in Cu-Cu collisions at \sNN$~=~0.2$~TeV and Au-Au collisions at \sNN$~=~39$ and $62.4$ GeV. 
\begin{figure}
 \includegraphics[height=0.196\textheight]{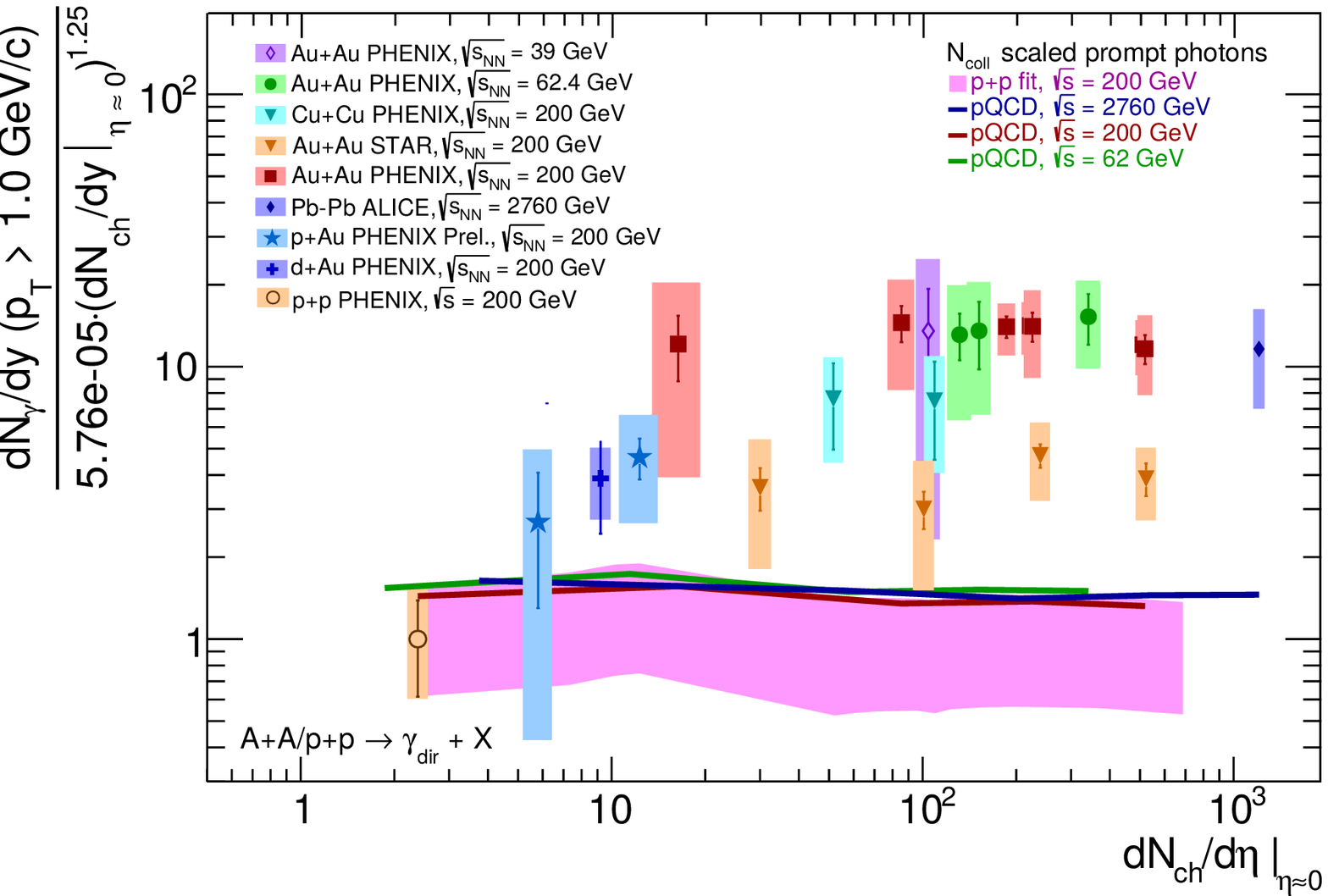}
 \includegraphics[height=0.196\textheight]{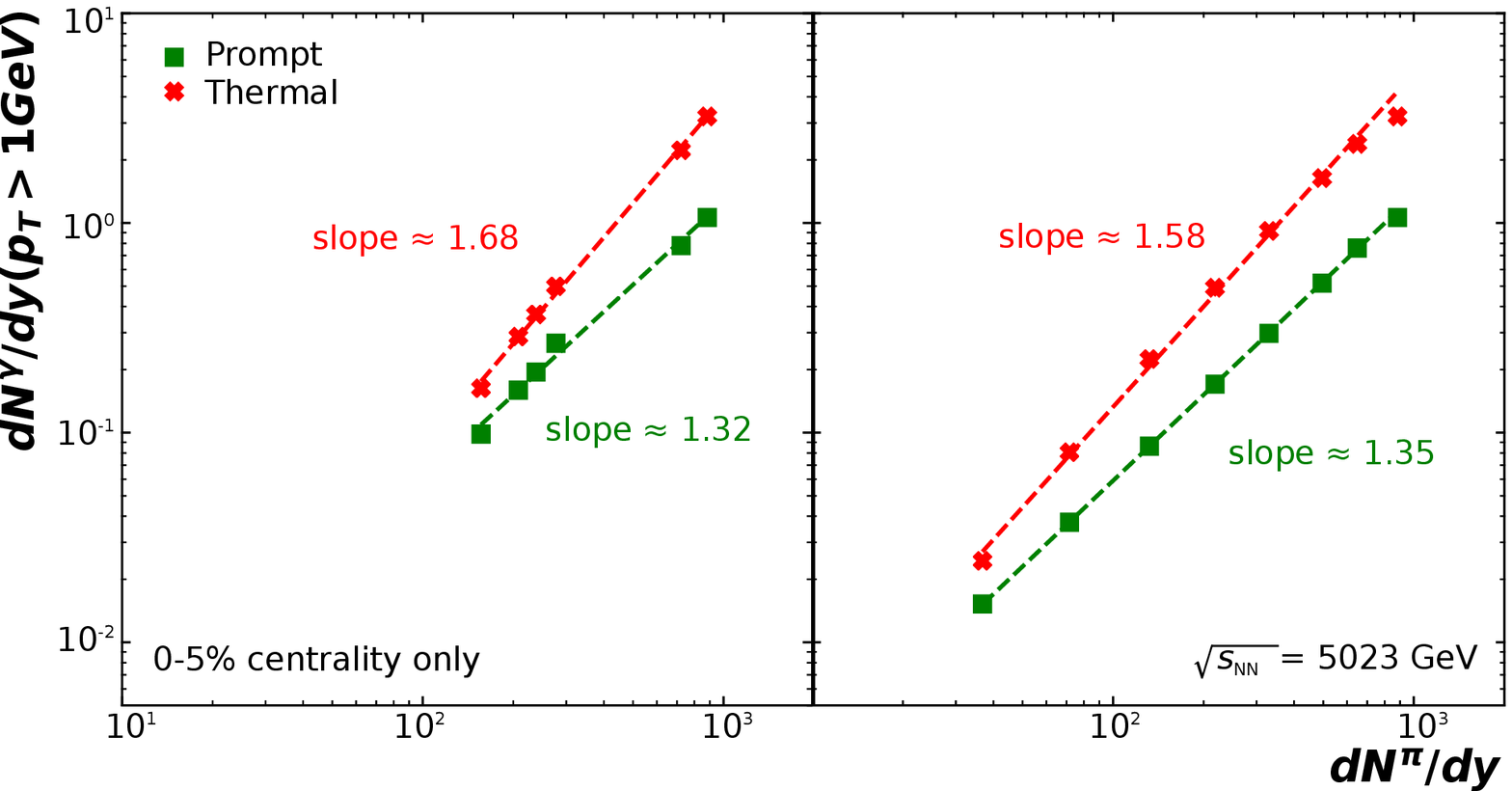}
 \caption{Left: Direct photon integrated yield for $\pt~>~1$ GeV/$c$ as a function of the number of charged particles (\Nch) at mid rapidity. The yield has been normalized to corresponding integrated photon yield in pp collisions at \spp$~=~0.2$ TeV scaled by the number of charged particles to the power of $\alpha =1.25$ \cite{Khachatryan:2018evz}. Middle: Integrated photon yield for prompt and thermal photons above 1 GeV/$c$ as function of \Nch\ for 0-5\% central events at different collision energies \cite{Gale:2019abf}. Right: Integrated photon yield for prompt and thermal photons above 1 GeV/$c$ as function of \Nch\ for different centralities in Pb-Pb collisions at $\sNN~=~5.02$ TeV \cite{Gale:2019abf}.}
 \label{fig:Scaling}
\end{figure}
The compilation of the \Ncoll-scalled direct photon spectra in A-A collisions for various species and centralities can be found in \ifarxiv \href{fig:dirGppandPbPb}{Figure \ref{fig:dirGppandPbPb} (right)} and \fi \href{fig:xTppandAA}{Figure \ref{fig:xTppandAA} (right)} as function of \xt. Once more it can be seen that \xt\ and \Ncoll-scaling holds at high transverse momenta within $20-50\%$. Taking into account only the lower transverse momentum points for the various collision energies they seem to follow a second \xt-scaling curve with a similar exponent. The PHENIX collaboration investigated this scaling property in more detail as a function of the number of charged particles at mid rapidity \cite{Adare:2018wgc} by integrating the extracted photon spectra above 1 GeV/$c$. \href{fig:Scaling}{Figure \ref{fig:Scaling} (left)} shows the obtained scaling relation including the integrated yields measured in pp and p-A collisions. The points are normalized to the expected scaling of the photons based on the observed charged particle scaling. Considering only the points from the PHENIX and ALICE collaborations the direct photon yields seems to rise with increasing number of charged particles and then saturates at a common value for Au-Au and Pb-Pb collisions at all center of mass energies. The onset and later saturation is interpreted by the PHENIX collaboration as an indication of the QGP being produced in a fraction of the events and, when it reaches the plateau, in all of the collisions \cite{Adare:2018wgc}. However, such an independence of the collision energy cannot be explained easily by theoretical calculations \cite{Gale:2019abf} as seen in \href{fig:Scaling}{Figure \ref{fig:Scaling} (middle)}, while it can be understood that the direct photon yield scales as function of charged particle multiplicity when only considering one collision system and energy as seen in \href{fig:Scaling}{Figure \ref{fig:Scaling} (right)}. Furthermore, the scaling relation is not seen as clearly when also including the corresponding integrated direct photon yields at \sNN$ = 0.2$~TeV measured by the STAR collaboration. Their results are significantly lower than those obtained by the PHENIX collaboration. Thus it cannot be determined conclusively whether these measurements indicate that a QGP can be create in high multiplicity p-A collisions without further measurements from the various experiments.\\
In addition to the direct photon spectra measurements the simultaneous development of direct photon flow in A-A collisions was discussed at the conference based on the new results obtained by the ALICE collaboration \cite{Acharya:2018bdy}. In contrast to the results obtained by the PHENIX collaboration \cite{Adare:2015lcd} the spectra and flow measurements can be described by the theoretical calculations within $1-2 \sigma$. However, the uncertainties at the LHC are significantly larger than those achieved by the PHENIX collaboration due to the smaller signal to background ratio in the thermal photon transverse momentum region.\\
Additionally, the first di-muon measurements from the STAR collaboration in pp and p-Au at $\sNN~=~0.2$ TeV at mid-rapidity have been presented using neural networks to cope with the low purity and statistics in the $\mu^+\mu^-$ channel at low invariant masses \cite{Brandenburg:2018lmmu}. While the di-muon measurement does not seem to be feasible in Au-Au collisions, the same techniques can be applied to the di-electron measurements to improve the results for the large data sets obtained during the beam energy scan and using isobaric ions. The ALICE collaboration was able to extract the di-electron spectra in pp collisions at $\spp = 7$ and $13$~TeV as well as in Pb-Pb collisions at $\sNN = 2.76$ and $5.02$ TeV \cite{Acharya:2018ohw,Acharya:2018kkj,Acharya:2018nxm,Caliva:2018lme}. The obtained Pb-Pb data are in good agreement with cocktail calculations without in medium modifications of the $\rho^0$ between $150 < \Mee < 700$ MeV/$c^2$. However, there is a strong indication that the charm suppression needs to be taken into account for the cocktail calculations at $\sNN~=~5.02$~TeV. From the pp measurements constraints on the correlation between the total $c\bar{c}$ and $b\bar{b}$ cross sections could be extracted. Furthermore, no enhanced di-electron production with respect to the cocktail calculations is seen in high multiplicity pp collisions at $\spp = 13~$TeV indicating that the possible QGP contributions should be small if present. \vspace{-0.3cm}

\section{Light-by-Light Scattering in Heavy Ion Collisions}
\label{sec:lbl}
\vspace{-0.3cm}
\begin{figure}[t]
 \centering
 \includegraphics[height=0.19\textheight]{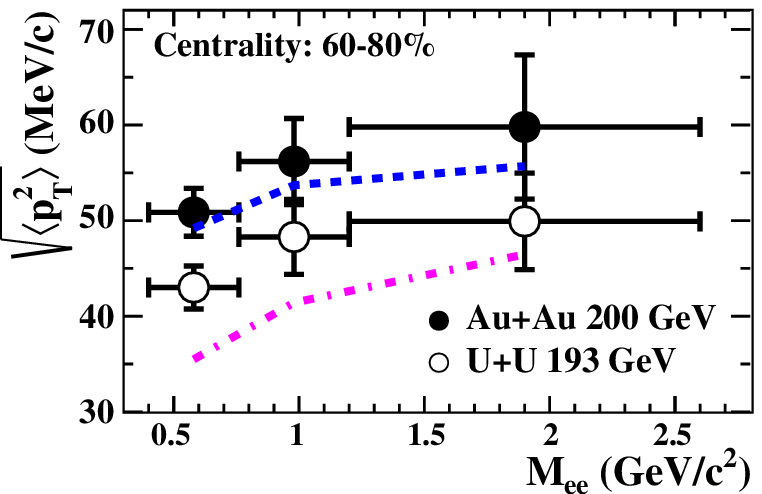} \hspace{0.1cm}
 \includegraphics[height=0.19\textheight]{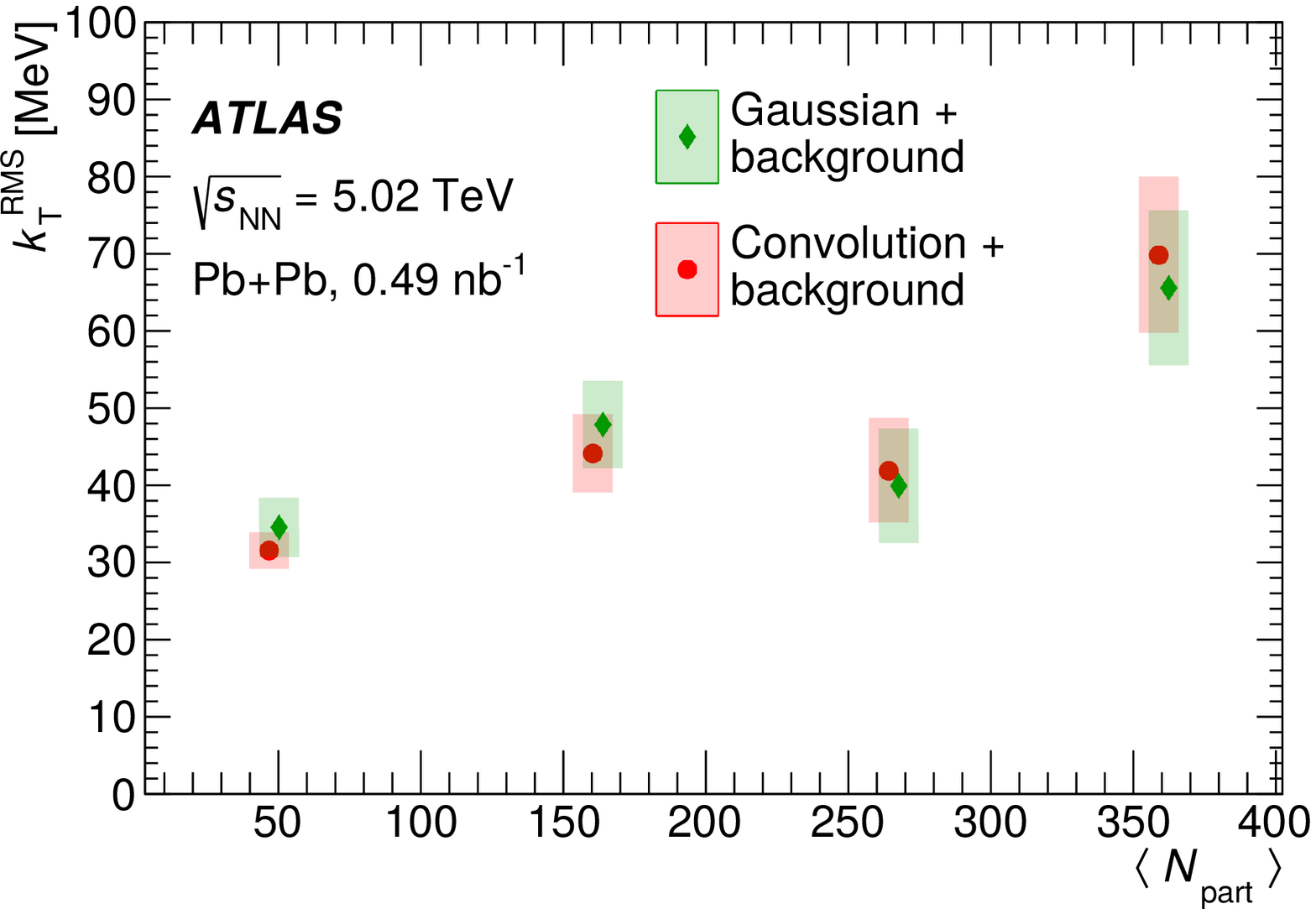}
 \caption{Left: Distribution of the $\sqrt{ \langle \pt^2 \rangle}$ of excess di-electron yields in 60-80\% peripheral Au-Au and U-U collisions measured by the STAR collaboration \cite{Adam:2018tdm}. Right: Mean transverse momentum broadening ($\kt^{RMS}$) obtained from Gaussian fits to the acoplanarity distribution of high momentum di-muon events in Pb-Pb collisions at $\sNN~=~5.02~$TeV by the ATLAS collaboration as a function of the number of participants (\Npart) \cite{Aaboud:2018eph}.}
 \label{fig:lbl}
\end{figure}
In heavy-ion collisions at RHIC and LHC energies strong magnetic fields are created by the passing nuclei at large distances from the nuclei cores. These fields generate large quasireal photon fluxes and can lead to light-by-light scattering. The first evidence of this process has been presented by the ATLAS collaboration with $4.4 \sigma$ \cite{Aaboud:2017bwk} and was now confirmed by the CMS collaboration with $4.1 \sigma$ \cite{Sirunyan:2018fhl} in the two photon final state in ultraperipheral Pb-Pb collisions at \sNN$~=~5.02$~TeV. The respective measured fiducial cross sections are $\sigma_{fid}~=~70 \pm 24$ (stat.) $\pm~17$(syst) nb and $\sigma_{fid}~=~120~\pm~46$ (stat) $\pm~4$ (theo) nb. Using these data the CMS collaboration was also able to improve the limits on the axion-like particle production in the $\gamma\gamma$-channel as no significant excess was observed in the di-photon invariant mass distribution \cite{Sirunyan:2018fhl}.\\
Light-by-light scattering can also occur in more central collisions, however, it is more difficult to observe as the resulting final state particles are observed in parallel to a large amount of particles produced in the actual ion collisions. The STAR collaboration presented results on the di-electron invariant mass spectra in peripheral events showing a significant excess in the low and intermediate mass region above the hadronic cocktail \cite{Adam:2018tdm}. The excess could be explained by the superposition of the light-by-light scattering into $e^+e^-$ in these events. In this case the electron pair \pt\ distribution is significantly enhanced at low \pt\ with respect to the expectations from the hadronic cocktail calculations. Furthermore, the \pt$^2$ distribution of the excess yield appears to be broadened beyond the expectations based on models describing the contributions from UPC events. The additional broadening and its invariant mass dependence can be described by introducing an additional strong magnetic field perpendicular to the beam line (blue dashed line in \href{fig:lbl}{Figure \ref{fig:lbl} (left)}). \\
The ATLAS collaboration, on the other hand, looked at di-muon events overlaid with the normal heavy-ion collision requiring the events to have at least two muons with single muon $\pt > 4$ GeV/$c$ and di-muon invariant masses of $4 < \Mmm < 45$ GeV/$c^2$ \cite{Aaboud:2018eph}. The pair acoplanarity ($\alpha$) and the asymmetry distributions ($A$) have been corrected for the background contributions from heavy flavor decays based on data driven template fits outside the signal region and were afterwards compared to the corresponding distributions obtained from ultraperipheral collisions using the same criteria. It can be seen that the $\alpha$ distribution broadens significantly when going to more central events. This can be interpreted as the muons being slightly deflected by the presence of an additional electromagnetic field obtaining a small amount of transverse momentum $|\kt| \ll \pt^\pm$. The corresponding $\kt^{RMS}$ values obtained from Gaussian fits to the $\alpha$ distributions after subtraction of the expected UPC background in different centrality intervals can be seen in \href{fig:lbl}{Figure \ref{fig:lbl} (right)}. In 0-10\% central collisions an average transverse momentum transfer of $70~\pm~10$ MeV/$c$ can be observed. \\
While the results from both collaboration indicate a deflection of the leptons by an electromagnetic source, it remains to be seen whether it will be possible to measure the field strength of such a source taking into account all other possible background contributions and its origin \cite{Klein:2018fmp}. \vspace{-0.3cm}

\section{Isolated Photons as Calibration and Tagging Objects for Jet Modification Studies in p-A and A-A Collisions}
\label{sec:probes}
\vspace{-0.3cm}
Through the study of the direct photons not only the QGP properties can be explored, but also a reference for distinguishing initial and final state effects seen in other probes can be established. 
The isolated prompt photons mainly produced in the initial hard scattering can be used as calibration and tagging probe for the recoiling jets. 
Those jets will mainly originate from quarks and can thus help to distinguish the effects of the final state interaction in the QGP for quarks and gluons. 
Furthermore, the energy of the photon serves as a measure of the initial energy of the jet, as the photon will leave the QGP unaffected by the medium, while the constituents of the quark-jet will interact with it. \\
At this conference the first results on isolated photon-hadron and jet correlations measured by the ALICE collaboration in pp and p-Pb collisions at \sNN$~=~5.02$ TeV have been presented for intermediate photon transverse momenta between $10$ and $40$ GeV/c \cite{Arratia:2018nra}. It has been found that the fragmentation functions do not seem to be modified in minimum bias p-Pb collisions, while it is expected that the strongest modifications for the recoiling jets in heavy-ion collisions at the LHC should be seen in this transverse momentum region. Similar observations for pp and p(d)-Au collisions at $\sNN~=~200$ GeV were presented by the STAR \cite{Sahoo:2018sph} and PHENIX \cite{Osborn:2018wxd} collaborations at slightly lower \pt\ for the isolated photons. Additionally, these collaborations presented the corresponding results in Au-Au collisions at the same \sNN, which strongly indicate a redistribution of the recoiling jet energy towards large angles. \\
\begin{figure}[t]
 \includegraphics[height=0.19\textheight]{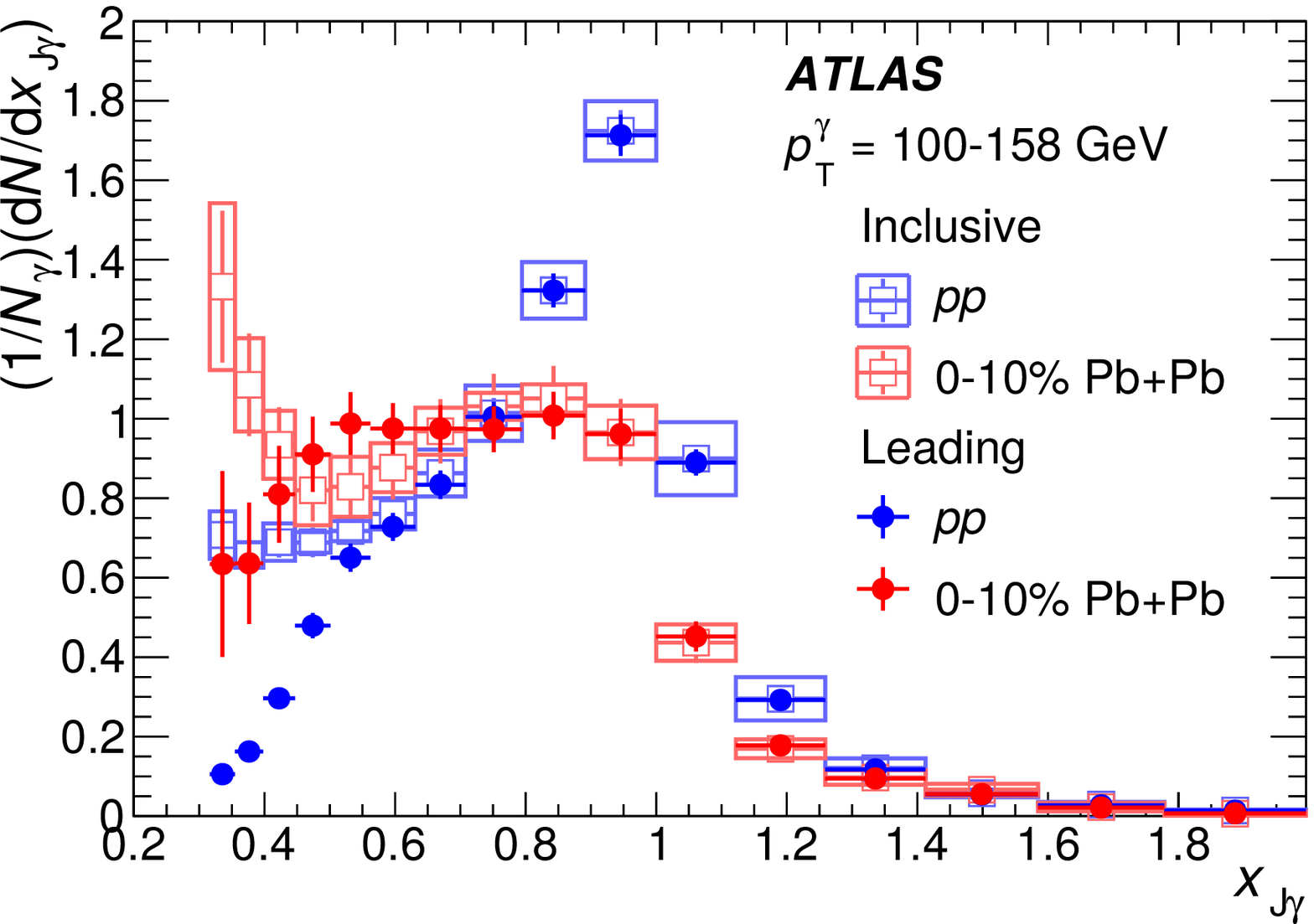}
 \includegraphics[height=0.20\textheight]{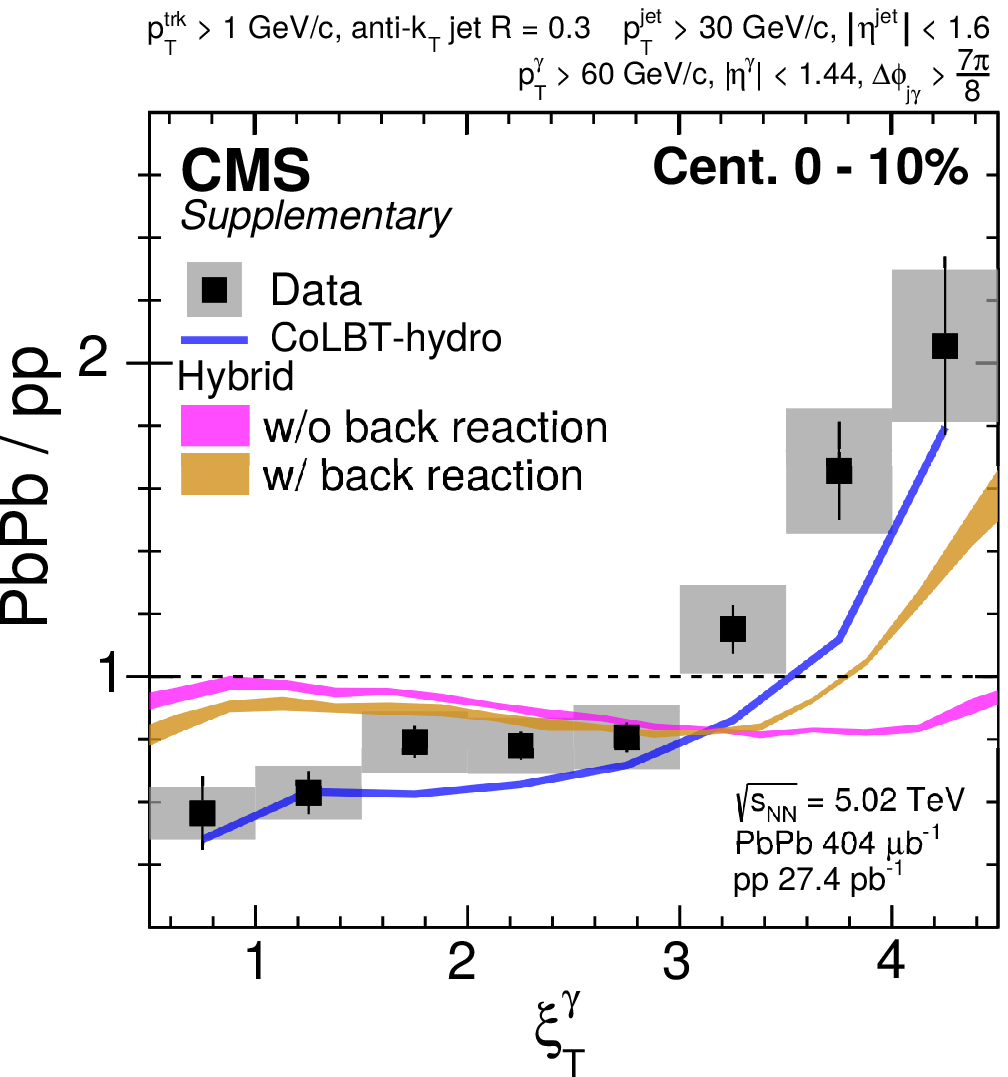}
 \includegraphics[height=0.20\textheight]{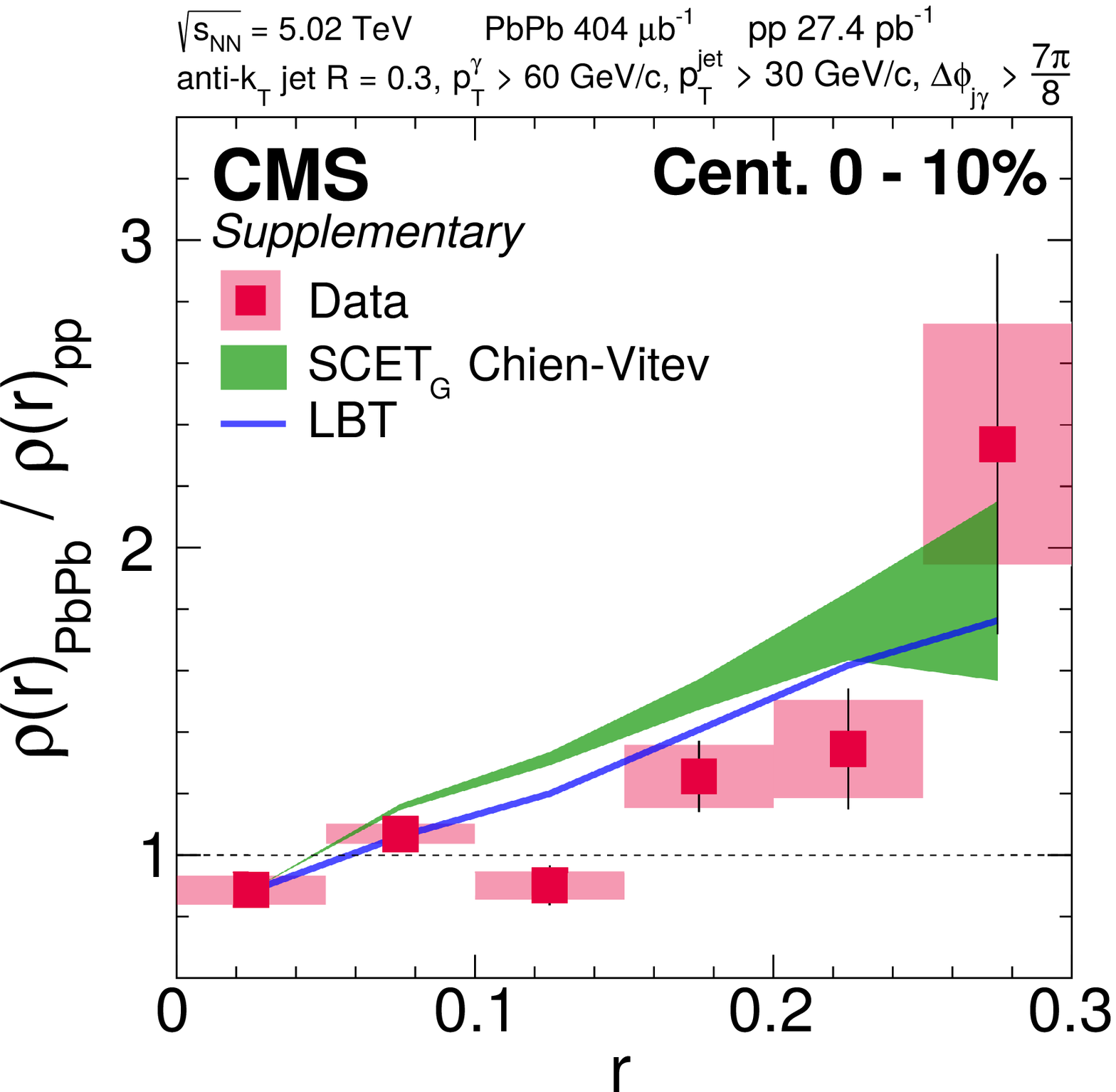}
 \caption{Left: Photon-jet \pt-balance distribution ($x_{J\gamma}$) in pp (blue) and 0-10\% central Pb-Pb collisions at $\sNN~=~5.02$ TeV for the inclusive jet (closed) and photon-plus-leading-jet (open) selection for $\pt^\gamma = 100-158$ GeV \cite{Aaboud:2018anc}. Middle: Ratio of the \xitg\ distribution in 0-10\% central Pb-Pb and pp collisions at $\sNN~=~5.02$ TeV compared to various theoretical models \cite{Sirunyan:2018qec}. Right: Ratio of the differential jet shape $\rho(r)$ for jets associated with an isolated photon with a \pt\ above $60$ GeV/c for 0-10\% central Pb-Pb and pp collisions at $\sNN~=~5.02$ TeV compared to different theory calculations \cite{Sirunyan:2018ncy}.}
 \label{fig:gammajet}
\end{figure}
Furthermore, the ATLAS and CMS collaborations presented their recent multi-differential results on the isolated-photon jet tagged measurements \cite{Aaboud:2018anc,Sirunyan:2017qhf,Sirunyan:2018qec,Sirunyan:2018ncy} at high \pt. 
When measuring the $\gamma$-jet \pt-balance distributions in multiple centrality classes the ATLAS collaboration observed a significant broadening of the $x_{j\gamma}$ distribution in central Pb-Pb events with an isolated photon and a recoiling leading jet (\href{fig:gammajet}{Figure \ref{fig:gammajet} (left)}). 
The broadening increases further when considering all jets in the $\gamma$-tagged events. 
Such a smearing can be explained by a strong variation in the jet-by-jet energy loss for instance depending on the path length in the medium of the recoiling quark. 
Furthermore, a clear difference in the evolution of $\gamma$-tagged jet fragmentation functions compared to those for inclusive jets has been found by the ATLAS collaboration going from pp to central Pb-Pb collisions \cite{ATLAS:2017isy}, which might be partially explained by a flavor depended energy loss of the jets. 
The CMS collaboration, on the other hand, presented their results regarding the isolated photon tagged jet shapes in pp and Pb-Pb collisions at $\sNN~=~5.02$ TeV \cite{Sirunyan:2018ncy}. 
The corresponding comparisons for pp and central Pb-Pb collisions are shown in \href{fig:gammajet}{Figure \ref{fig:gammajet}} as function of \xitg\ (middle) and $r$ (right) for photons with a minimum \pt\ of 60 GeV/$c$. 
These indicate an enhancement of low \pt\ particles and a depletion of high \pt\ particles inside the jet, which is stronger for $\gamma$-tagged jets than for inclusive jets.
Furthermore, it can be seen that the jet momentum is distributed at greater radial distance in Pb-Pb collisions, which can be interpreted as a direct observation of jet broadening in the QGP. 
By using the $\gamma$-tagged jets the energy scale of the jet is fixed which eases the comparisons to theory calculations, which can qualitatively reproduce the distributions, but still require some fine tuning in terms of the magnitude of the effects seen in Pb-Pb collisions. \vspace{-0.3cm}

\section{Photons and Bosons as Probes for the Initial State and Scaling Properties}
\label{sec:IS}
\begin{figure}[t]
  \centering
 \includegraphics[height=0.259\textheight]{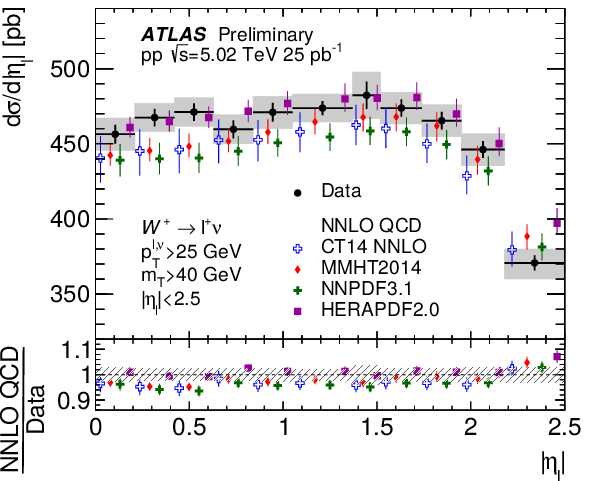} \hspace{0.1cm}
 \includegraphics[height=0.28\textheight]{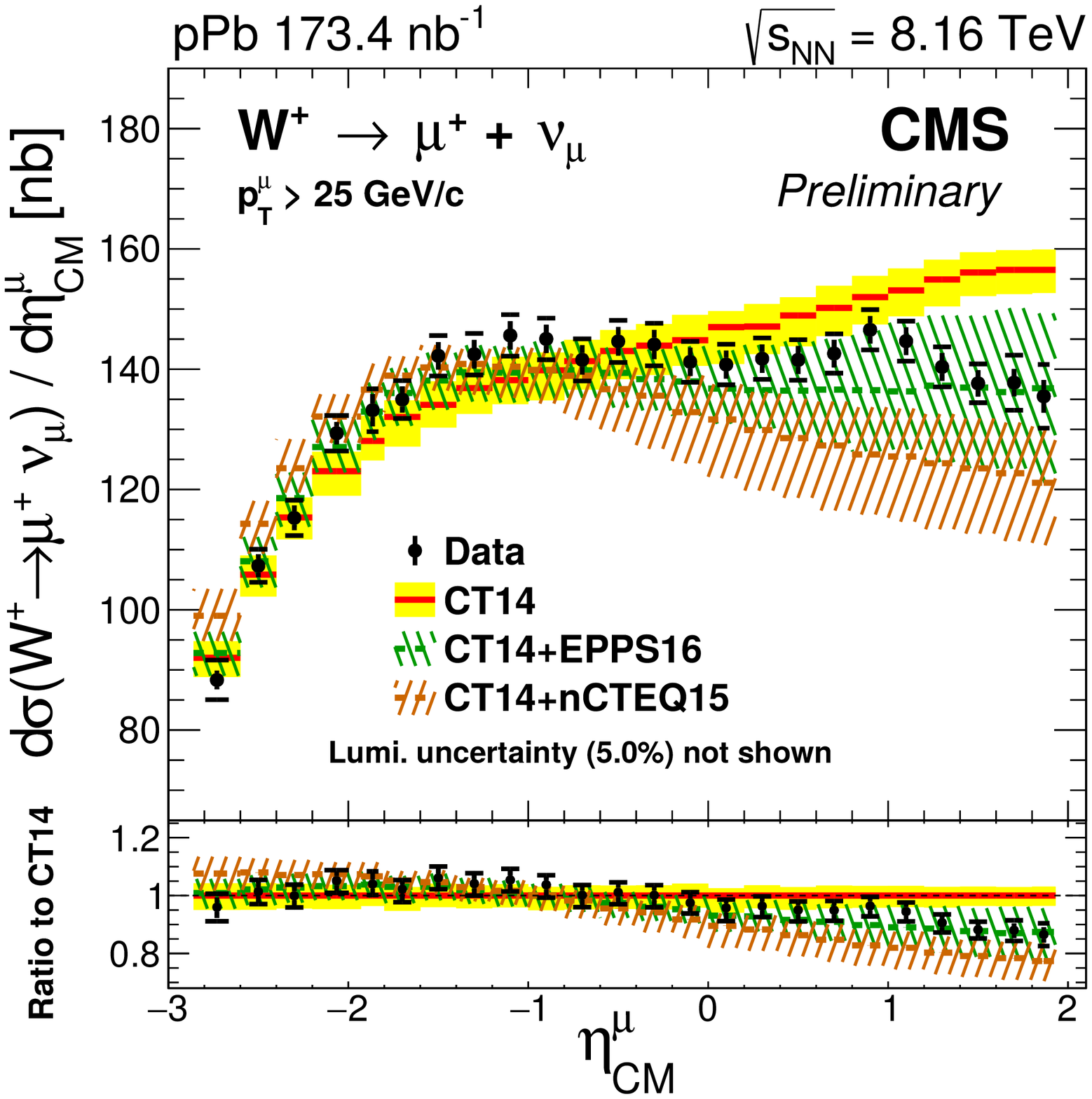}
 
 \caption{Left: Differential W$^+ \rightarrow \mu^+ + \nu_\mu$ production cross section as a function of the lepton pseudo-rapidity in pp collisions at $\spp~=~5.02$ TeV compared to different NNLO pQCD calculations with different PDFs as input \cite{Aaboud:2018nic}. Right: W$^+ \rightarrow \mu^+ + \nu_\mu$ production cross section as a function of the muon pseudo rapidity in p-Pb collisions at $\sNN~=~8.16$ TeV \cite{CMS:2018ilq}.}
 \label{fig:Ws}
\end{figure}

\vspace{-0.3cm}
As the electroweak probes at high \pt\ and masses at mid-rapidity are not modified by the QGP, they can also be used in order to calibrate the Glauber model calculations and their estimated \Ncoll\ in particular in p(d)-A collisions, where the correlation between \Ncoll\ and \Nch\ is not that strong. At higher rapidities the photons, W$^\pm$ and Z$^0$ can help to constrain the nuclear parton distribution function (nPDFs). At this conference recent results on the high \pt\ isolated photon \cite{ATLAS:2017ojy} and W$^\pm$ \cite{CMS:2018ilq} production in p-Pb collisions at $\sNN~=~8.16$ TeV  have been discussed in different rapidity intervals. While the isolated photons can only mildly discriminate between different nPDFs within the current uncertainties. Stronger constrains for the nPDFs can be derived from the W$^\pm$, which favor the nPDFs from \cite{Eskola:2016oht} as can be seen in \href{fig:Ws}{Figure \ref{fig:Ws} (right)}. Additionally, the pp measurements of the weak bosons have been improved \cite{Aaboud:2018nic} and consequently an updated nuclear suppression factor (\RAA) could be obtained with smaller uncertainties than those of the Glauber model. Thus it might now be possible to replace the traditional \RAA\ measurements for hard probes by $\ZAA~=~(\NAA^X \cdot \sipp^Z) / (\NAA^Z \cdot \sipp^X)$ measurements, which do not depend on model calculations any longer. The multi-differential measurements of the W$^\pm$ production can also be used to test and constrain the existing PDFs for the proton as well as NNLO pQCD calculations for the weak boson production in pp collisions. The comparisons of the most recent NNLO pQCD calculations to the measured W$^+$ production cross section in pp collisions at $\spp~=~5.02$ TeV  as a function of the lepton pseudo-rapidity can be seen in \href{fig:Ws}{Figure \ref{fig:Ws} (left)}. It can be seen that the best agreement can be reached when using the HERAPDF2.0 PDF sets, while the remaining calculations slightly underestimate the cross sections in nearly all pseudo-rapidity bins.

\ifarxiv
  \bibliographystyle{utphys}
  \bibliography{Bibliography}{}
\else
  \input{bib_ordered2.tex} 
\fi

\ifarxiv
\section*{Appendix}
\begin{figure}
  \includegraphics[width=0.5\textwidth]{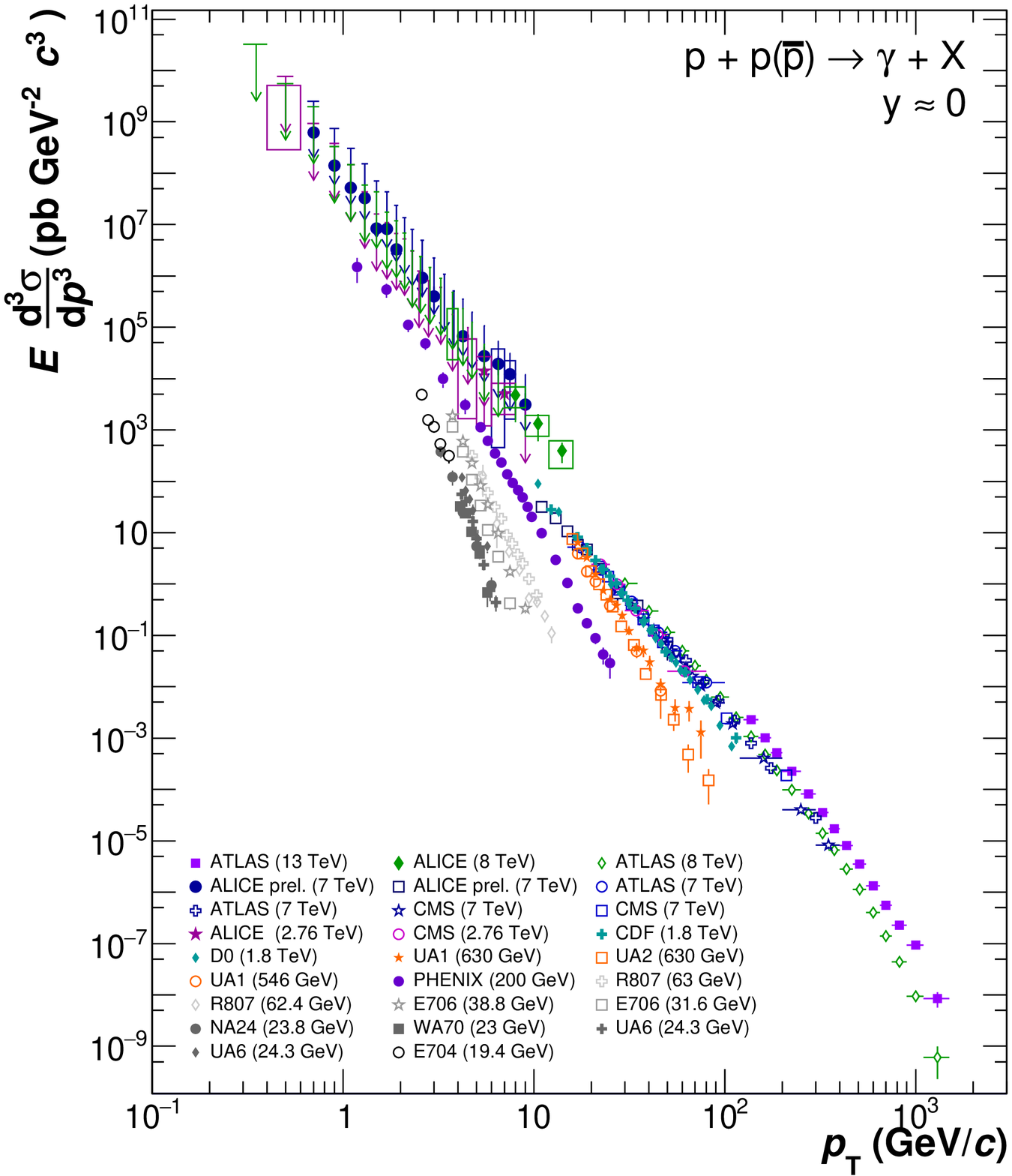} 
  \includegraphics[width=0.5\textwidth]{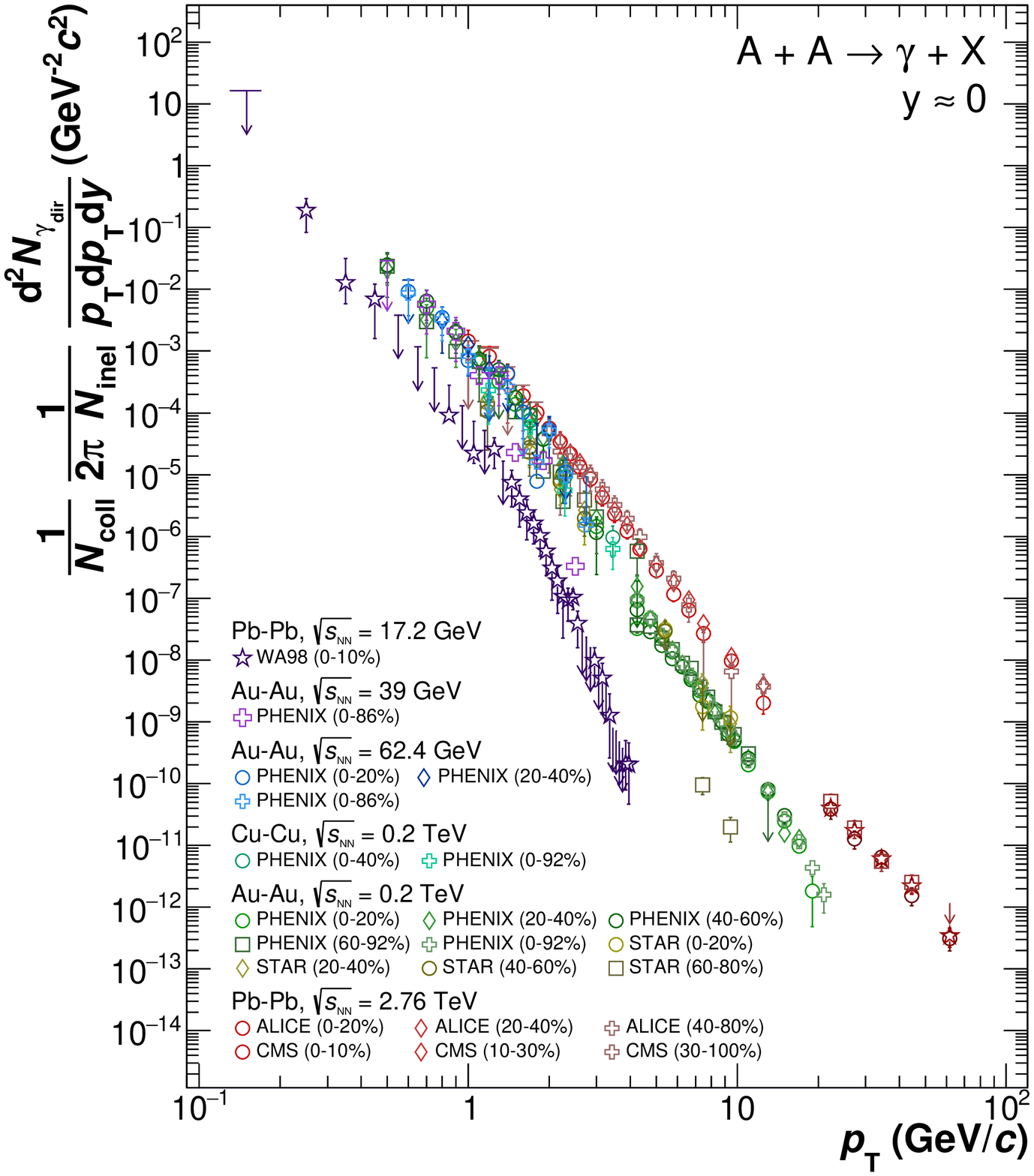}  
  \caption{Left: Summary of (isolated) direct photon invariant cross sections measured in pp and p$\bar{\mathrm{p}}$ collisions at various collider energies 
  \cite{Adare:2012yt,Abachi:1996qz,Abe:1994rra,Albajar:1988im,Albajar:1989sr,Alitti:1991yk,Adams:1995gg,Apanasevich:2005gs,Apanasevich:2004dr,
  Alverson:1993da,DeMarzo:1986vi,Bonesini:1987bv,Sozzi:1993sm,Ballocchi:1998au,Aad:2011tw,Aad:2010sp,Aad:2016xcr,Aaboud:2017cbm,Wilde:2012wc,Acharya:2018dqe,Chatrchyan:2011ue,Khachatryan:2010fm}. Right: Summary of \Ncoll-scaled (isolated) direct photon invariant yields measured in A-B collisions at various collider energies \cite{Gale:2009qy,Aggarwal:2000th,STAR:2016use,Adare:2018wgc,Adare:2018jsz,Adare:2014fwh,Afanasiev:2012dg,Adare:2009qk,Adare:2008ab,Adler:2005ig,Aad:2015lcb,Chatrchyan:2012vq,Adam:2015lda}. The yields from the different experiments have been scaled by \Ncoll\ obtained by the corresponding collaboration for the respective centrality interval.}
  \label{fig:dirGppandPbPb}
\end{figure}
\begin{figure}
  \includegraphics[width=0.5\textwidth]{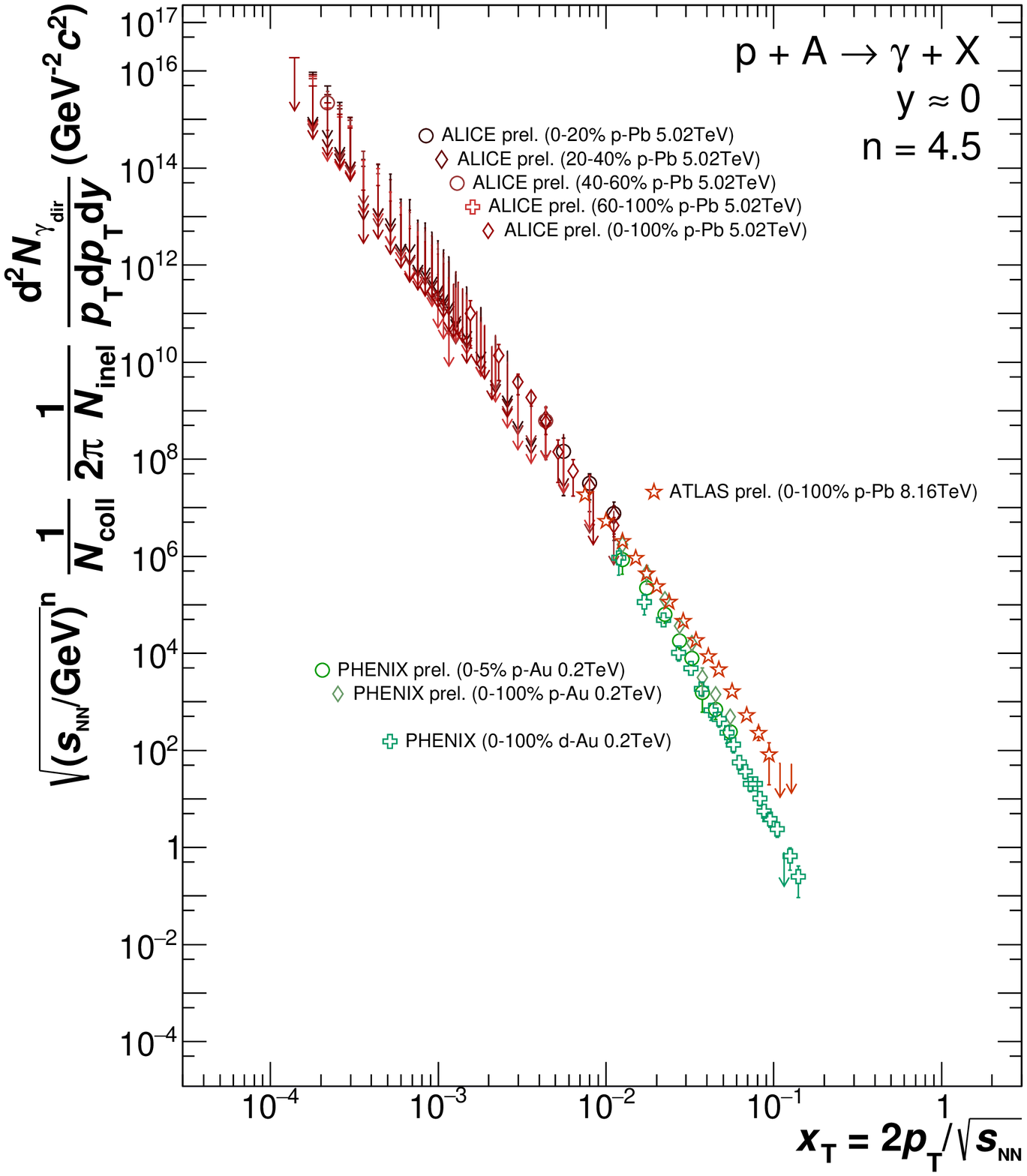} 
  \includegraphics[width=0.5\textwidth]{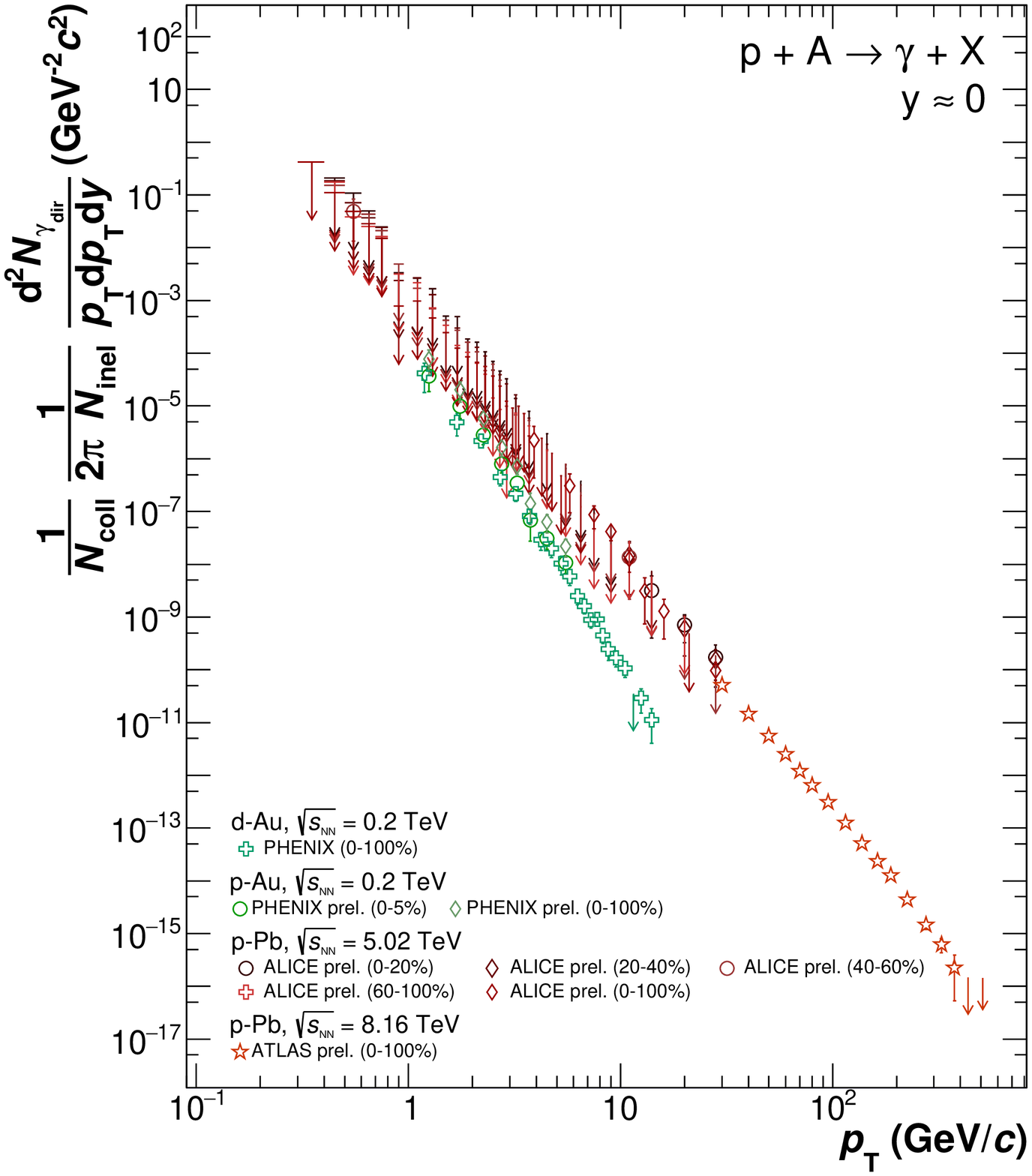} 
  \caption{Left: Summary of \Ncoll-scaled (isolated) direct photon invariant yields measured in p(d)-B collisions at various collider energies \cite{Adare:2012vn,Khachatryan:2018evz,Schmidt:2018ivl}. The yields from the different experiments have been scaled by \Ncoll\ obtained by the corresponding collaboration for the respective multiplicity interval. Right: Summary of \Ncoll\ and \xt-scaled (isolated) direct photon invariant yields measured in p(d)-B collisions at various collider energies using a common $n = 4.5$. }
  \label{fig:dirGpA}
\end{figure}
\begin{figure}
 \includegraphics[width=0.5\textwidth]{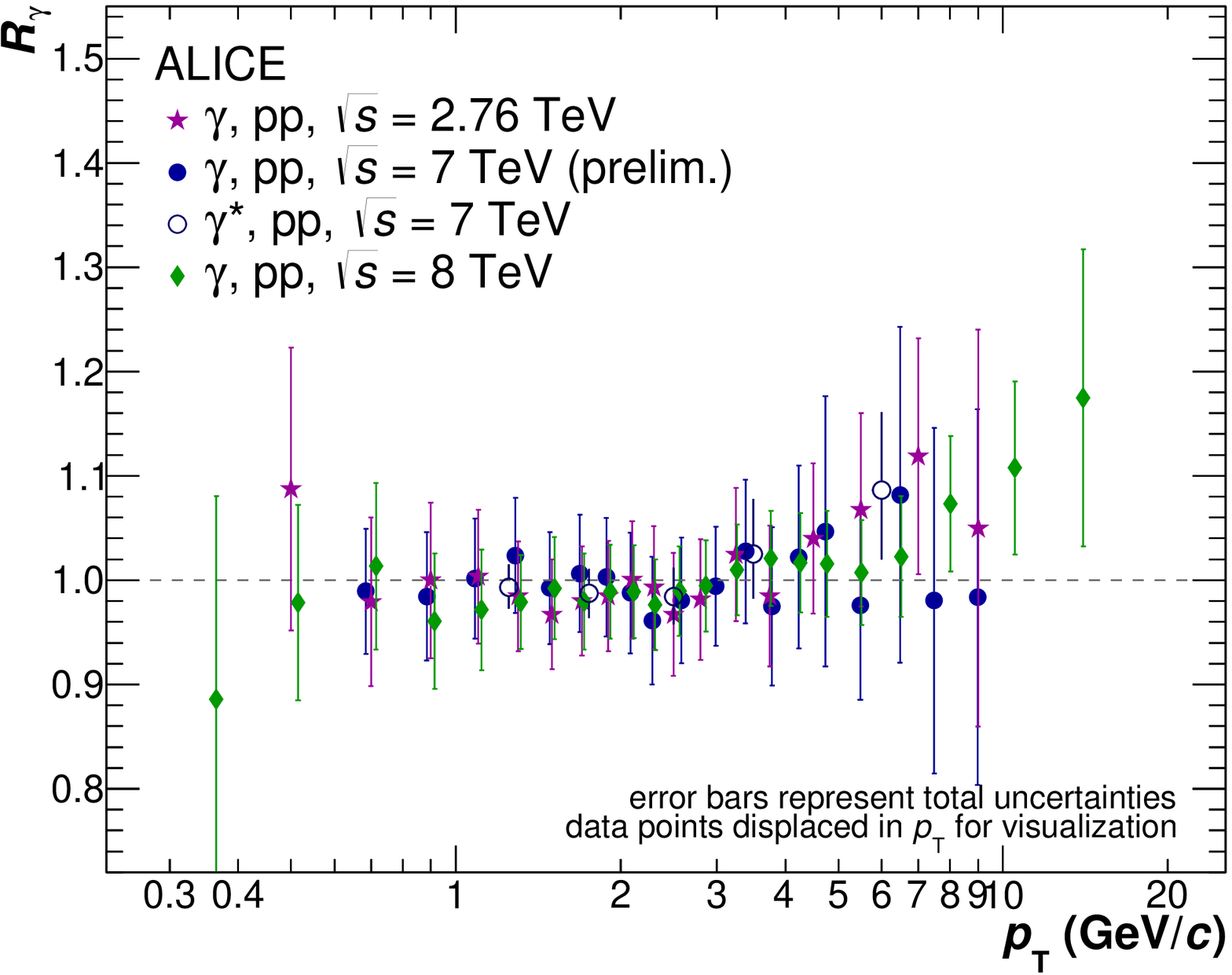}
 \includegraphics[width=0.5\textwidth]{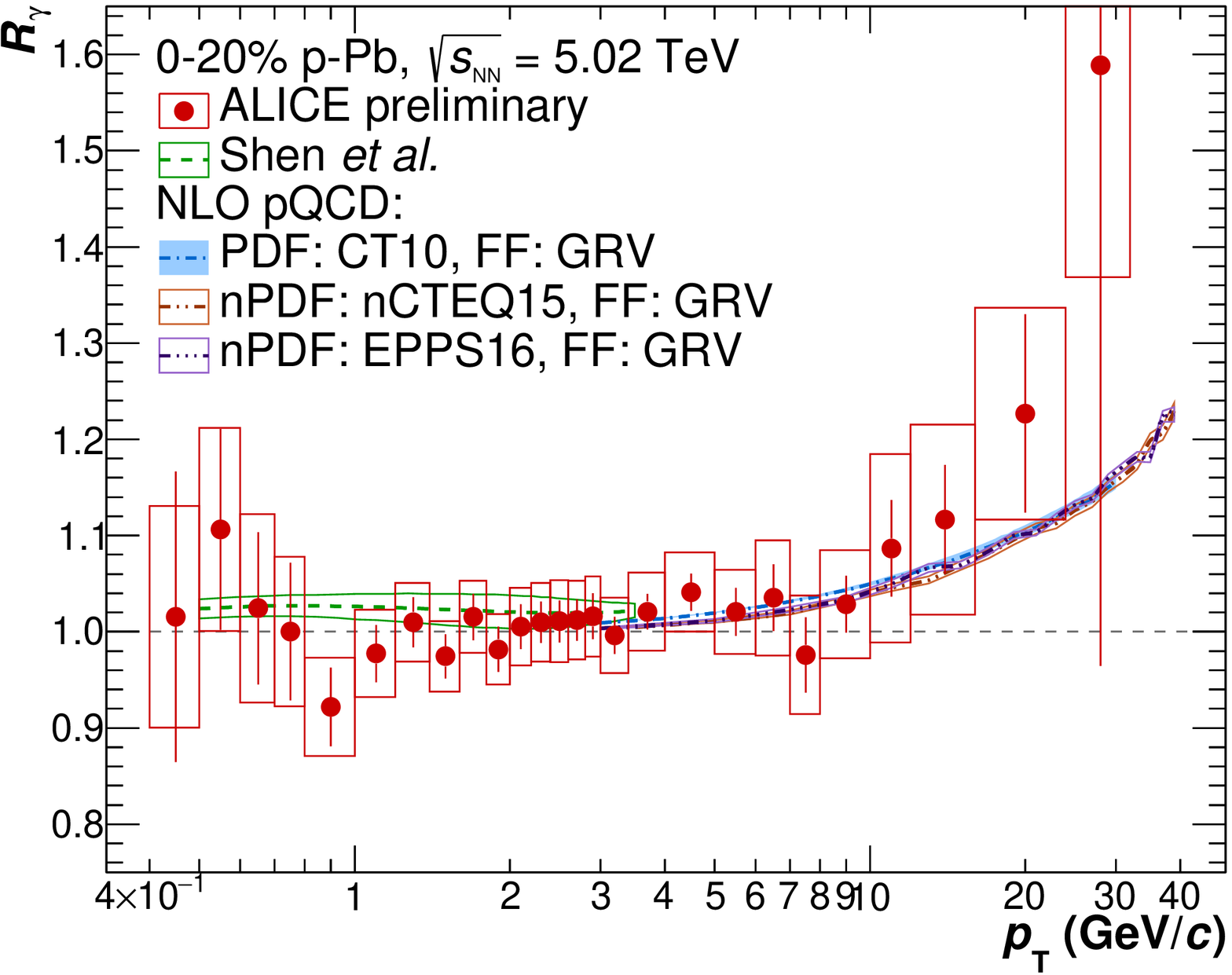}
 \caption{Left: Compilation of the direct photon excess ratio (\Rg) measurements provided by the ALICE collaboration in pp collisions at $\spp~=~2.76, 7$ and $8$ TeV \cite{Wilde:2012wc,Acharya:2018dqe,Acharya:2018ohw}. The vertical errors bars represent the total uncertainties. Right: Direct photon excess ratio in the 0-20\% most central p-Pb collisions at $\sNN~=~5.02$ TeV measured by the ALICE collaboration \cite{Schmidt:2018ivl} together with the most recent predictions from NLO pQCD and hydrodynamic calculations \cite{Shen:2016zpp}.}
 \label{fig:DRppandpPb}
\end{figure}

\fi

\end{document}